\documentclass[aps,prb,longbibliography,reprint]{revtex4-1}

\usepackage{graphicx}
\usepackage{bm}
\usepackage{bbold}
\usepackage{amssymb}
\usepackage{amsmath}
\newcommand{\red}[1]{{#1}}

\renewcommand{\vec}[1]{{\mathbf #1}}

\bibpunct{[}{]}{,}{n}{}{}

\begin{document}

\title{Localization of light in a three-dimensional disordered crystal of atoms}

\author{S.E. Skipetrov}
\email[]{Sergey.Skipetrov@lpmmc.cnrs.fr}
\affiliation{Univ. Grenoble Alpes, CNRS, LPMMC, 38000 Grenoble, France}

\date{\today}

\begin{abstract}
We demonstrate that a weak disorder in atomic positions introduces spatially localized optical modes in a dense three-dimensional ensemble of immobile two-level atoms arranged in a diamond lattice and coupled by the electromagnetic field. The frequencies of the localized modes concentrate near band edges of the unperturbed lattice. Finite-size scaling analysis of the percentiles of Thouless conductance reveals two mobility edges and yields an estimation $\nu = 0.8$--1.1 for the critical exponent of the localization length. The localized modes disappear when the disorder becomes too strong and the system starts to resemble a fully disordered one where all modes are extended.
\end{abstract}

\maketitle


\section{Introduction}

Photonic crystals are periodic arrangements of scattering units (typically, dielectric spheres or rods) that exhibit frequency ranges (band gaps) for which no optical modes exist in the infinite structure and light propagation is forbidden \cite{yablonovitch93,joan08}. Thus, photonic crystals play the same role for light as semiconductor crystals do for electrons. They have numerous promising prospects for applications in optical technologies and, in particular, for guiding of light \cite{joan97,russell03}, lasing \cite{painter99,park04} and quantum optics \cite{lodahl04,faraon08}.

Photonic crystals exist in nature \cite{vigneron12} (e.g., natural opals \cite{sanders64} or wings of some butterflies \cite{argyros02,remo16}) or can be fabricated using modern nanofabrication techniques \cite{wijnhoven98,blanco00,campbell00,marichy16}. However, neither nature nor humans do a perfect job and real-life photonic crystals always have some degree of imperfection: fluctuating sizes or positions of elementary building units, vacancies, interstitial or substitution impurities, cracks \cite{koenderink05,toninelli08}. Whereas these imperfections do not destroy the band gap provided that they are not too strong, they introduce an interesting new feature in the spectrum: spatially localized optical modes appear in the band gap, especially near its edges \cite{john87}. Localization of eigenmodes of wave equations or of eigenstates of the Schr\"{o}dinger equation by disorder is a ubiquitous phenomenon discovered by Philip Anderson \cite{anderson58} and bearing his name \cite{lagendijk09,segev13}. Anderson localization of electromagnetic waves in general and of light in particular has been predicted by Anderson himself \cite{anderson85} and by Sajeev John \cite{john84}. Later on, it has been observed in fully disordered one- \cite{berry97,kemp16}, quasi-one- \cite{chabanov00,chabanov01} and two-dimensional \cite{dalichaouch91,schwartz07} disordered media whereas observing it in three dimensions (3D) turned out to be difficult \cite{sperling16,skipetrov16njp}. Even though Sajeev John proposed a way to facilitate localization of light in 3D by using disordered photonic crystals instead of fully disordered suspensions or powders a long time ago \cite{john87,john91}, no clear experimental realization of this idea has been reported up to date. Some signatures of Anderson localization have been observed in reflection of short optical pulses from a disordered photonic crystal \cite{douglass11} although the authors did not claim the observation of Anderson localization.

The idea of facilitating localization of light in 3D by using a photonic structure with a band gap arises from the localization criterion following from the scaling \cite{abrahams79} and the self-consistent \cite{vollhardt80,vollhardt92} theories of localization \cite{john93}:
\begin{eqnarray}
{\cal N}_{\textrm{EM}}(\omega) D_0(\omega) \ell_0^*(\omega) \lesssim \red{\text{const} \sim 1},
\label{loccrit}
\end{eqnarray}
where ${\cal N}_{\textrm{EM}}(\omega)$ is the density of electromagnetic modes (states), $D_0(\omega) = \red{v_{\mathrm{E}} \ell_0^*(\omega)}/3$ is the \red{``bare'' diffusion coefficient of light (i.e., the value that the diffusion coefficient would have in the absence of localization effects)}, \red{$v_{\mathrm{E}}$ is the energy transport velocity \cite{albada91,tiggelen92}}, and \red{$\ell_0^*(\omega)$} is the \red{transport} mean free path \red{in the absence of localization effects}. \red{In a fully disordered isotropic medium without any short- or long-range order,  ${\cal N}_{\textrm{EM}}(\omega) \sim k(\omega)^2/v_{\mathrm{E}}$ and} we obtain the standard Ioffe-Regel criterion of localization: $\red{k\ell_0^*} \sim k\ell \lesssim \red{\text{const} \sim 1}$, where \red{$k(\omega)$ is the effective wave number, $\ell$ is the scattering mean free path, and we made use of the fact that $\ell_0^*$ and $\ell$ are of the same order}. This criterion corresponds to a very strong scattering with $\ell$ shorter than the wavelength of light. If, however, the density of states  ${\cal N}_{\textrm{EM}}(\omega)$ is suppressed with respect to its value in the \red{fully disordered} medium, the criterion (\ref{loccrit}) becomes easier to obey. In a photonic crystal, ${\cal N}_{\textrm{EM}}(\omega) \to 0$ near a band edge and hence localized states are expected to appear for arbitrary weak disorder \cite{john91}.

Large and dense ensembles of cold atoms constitute a new experimental platform for the investigation of multiple light scattering \cite{labeyrie99,labeyrie03,kaiser05}. The very good knowledge of the properties of individual, isolated atoms and the constantly increasing degree of control of large atomic ensembles make atomic systems ideal candidates for verifying the existing theoretical predictions as well as for going beyond them by playing the role of ``quantum simulators'' \cite{lewenstein07,sanchez10}. However, whereas Anderson localization of matter waves in 3D random optical potentials has been successfully realized \cite{jendr12,semeghini15}, the somewhat reciprocal situation of light localization by scattering on cold atoms turns out to be difficult to implement \cite{kaiser09}. In addition to experimental difficulties of producing cold atomic clouds that are large and dense at the same time, theoretical calculations have pointed out that the vectorial nature of electromagnetic waves and the dipole-dipole interaction between nearby atoms may be a fundamental obstacle for Anderson localization of light \cite{skip14prl,bellando14}. Applying a static magnetic field to suppress the dipole-dipole interactions is a possible way to circumvent this obstacle \cite{skip15prl,skip18prl} but strong fields are required \cite{skip16pra}. An easier way towards light localization by cold atoms may be to arrange atoms in a periodic 3D lattice and enjoy the relaxation of the localization criterion (\ref{loccrit}) near an edge of a photonic band gap.

In this paper, we investigate spatially localized quasimodes that are introduced in an open 3D diamond atomic lattice of finite size by a randomness in atomic positions. Randomly displacing the atoms from their positions in the lattice is different from introducing disorder by randomly removing the atoms---a situation studied in Ref.\ \onlinecite{antezza13}---and allows for varying the strength of disorder while keeping the atom number constant. Thus, we can follow a transition from the perfect photonic crystal for vanishing disorder to a fully disordered system for strong disorder. After discussing the impact of boundary states, we establish that for a moderate amount of disorder $W$, two localization transitions exist near edges of a photonic band gap that the diamond lattice exhibits. A finite-size scaling analysis of one of these transitions yields the precise position of the mobility edge and an estimation of the critical exponent $\nu$ of the localization length. Increasing $W$ eventually leads to the closing of the band gap and the disappearance of localized states. A relation between the band gap formation, Anderson localization, and the near-field dipole-dipole coupling between the atoms is conjectured. Finally, implications of our results to experiments with cold atoms are discussed.

\section{The model}
\label{sec:model}

\begin{figure}[t]
\includegraphics[width=\columnwidth]{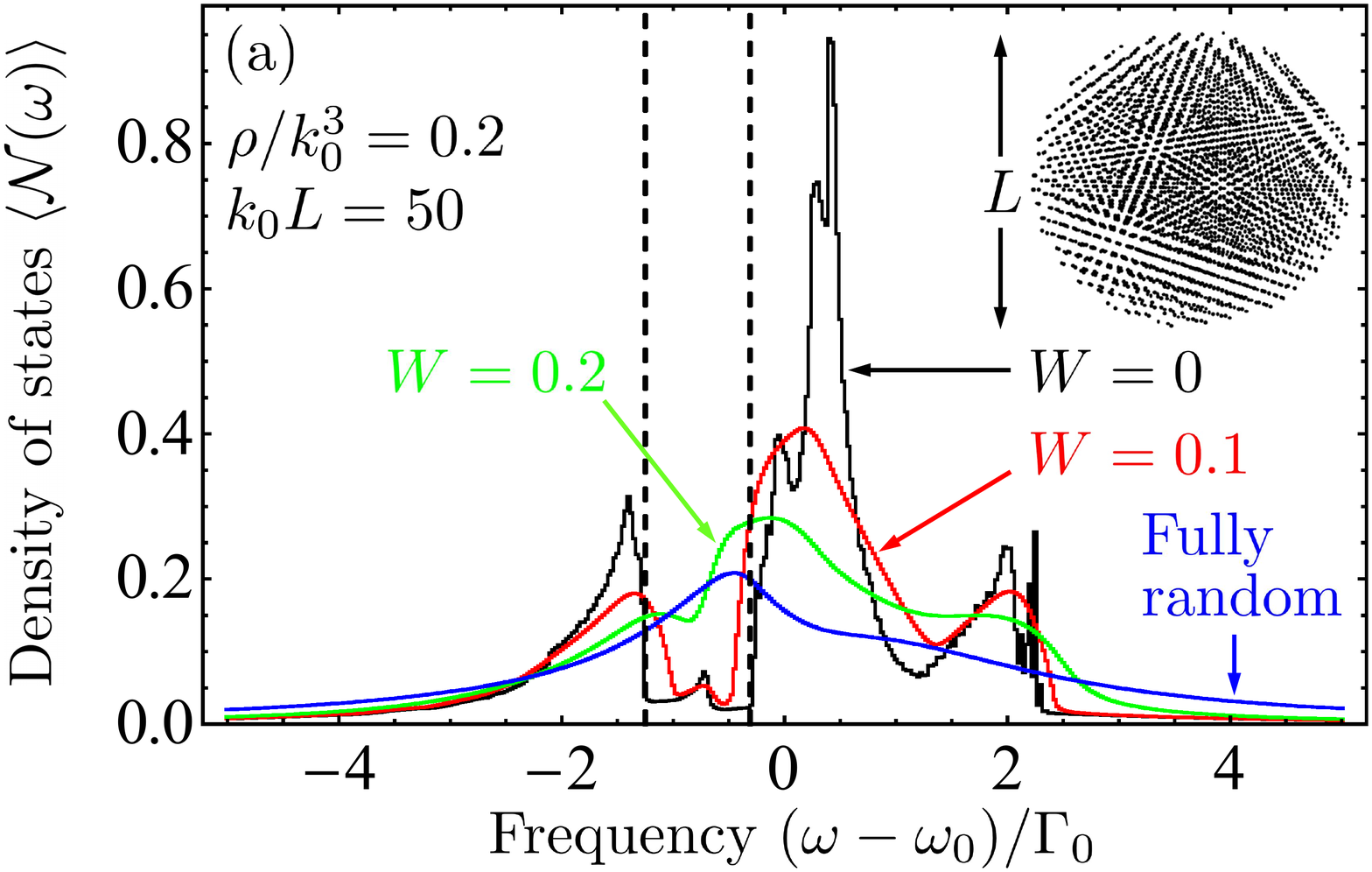}\\
\vspace{-6mm}
\includegraphics[width=\columnwidth]{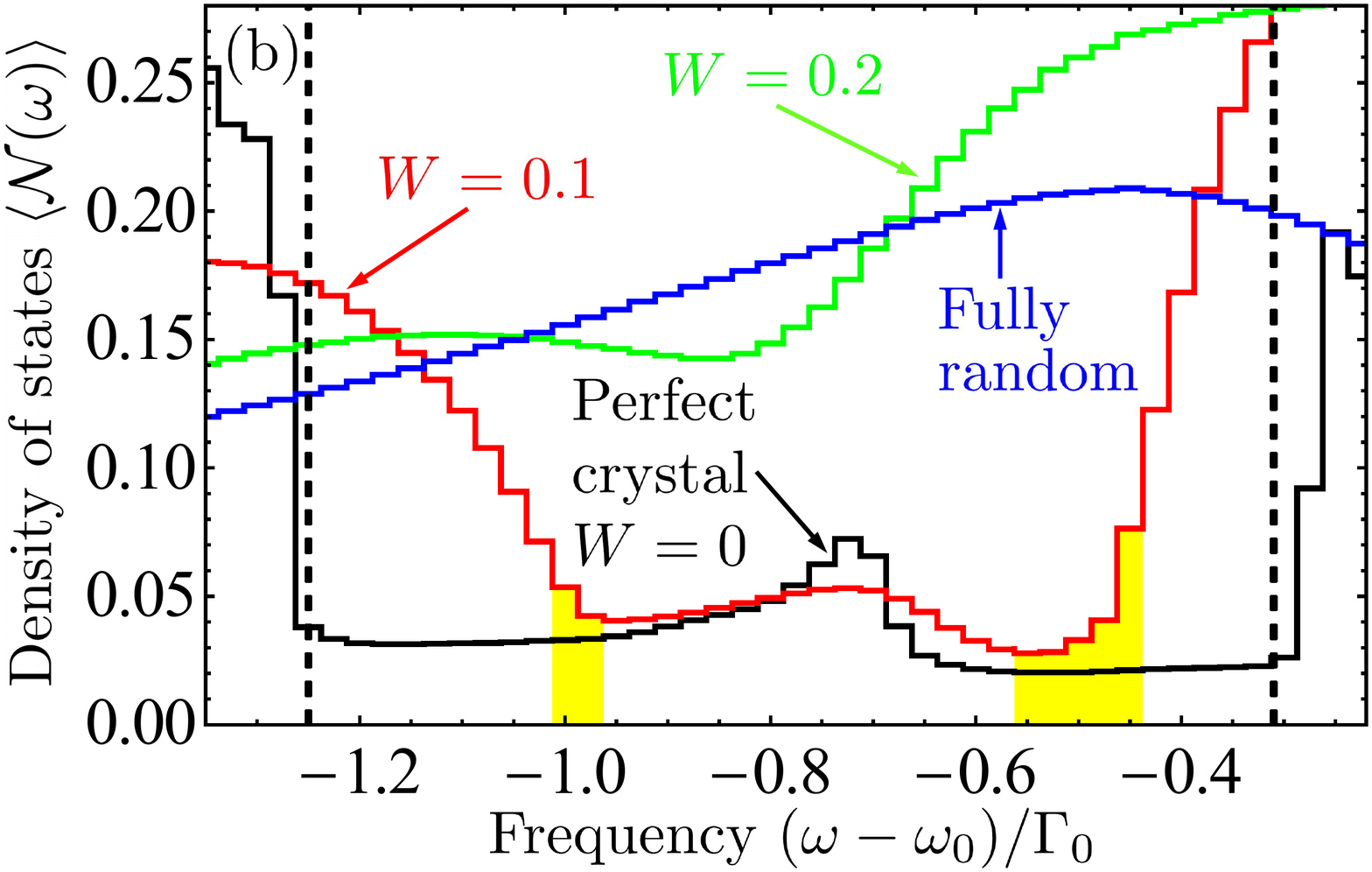}\\
\vspace{-8mm}
\caption{\label{fig_dos}
(a) Density of states of perfect ($W = 0$, black) and disordered (red, green, blue) photonic crystals for different disorder strengths: $W = 0.1$ (red), 0.2 (green). The blue line corresponds to a fully random ensemble of atoms. Averaging is performed over \red{461}, 175 and 82 random configurations for $W = 0.1$, 0.2, and the fully random case, respectively. Vertical dashed lines show band edges. Inset: A 3D rendering of a perfect diamond lattice of atoms. (b) Zoom on the band gap. Yellow shading shows frequency ranges in which we find localized quasimodes for $W = 0.1$.
}
\end{figure}

We consider \red{$N$} identical two-level atoms arranged in a diamond lattice. The lattice is a superposition of two face-centered cubic lattices (lattice constant $a$) with basis vectors $\vec{e}_1 = (0, a/2, a/2)$,  $\vec{e}_2 = (a/2, 0, a/2)$, $\vec{e}_3 = (a/2, a/2, 0)$ and $\vec{e}_1 + \vec{e}$, $\vec{e}_2 + \vec{e}$, $\vec{e}_3 + \vec{e}$, where $\vec{e} = (a/4, a/4, a/4)$. A sample of finite size is obtained from the unbounded lattice by keeping only the atoms inside a sphere of diameter $L$ \red{and volume $V = (\pi/6) L^3$} centered at the origin (see the inset of Fig.\ \ref{fig_dos} for a 3D rendering of the resulting sample). Disorder is introduced by displacing each atom by a random distance $\in [0, Wa]$ in a random direction, with $W$ being a dimensionless parameter characterizing the strength of disorder. The atoms have resonance frequencies $\omega_0$ and resonance widths $\Gamma_0$; their ground states have the total angular momentum $J_g = 0$ while their excited states have $J_e = 1$ and are thus three-fold degenerate, with the three excited states having the same energies but different projections $J_z = m$ ($m = 0$, $\pm 1$) of $\vec{J}_e$ on the quantization axis $z$. We have already used such a model of resonant two-level atoms coupled via the electromagnetic field to study random ensembles of atoms in our previous work \cite{skip14prl} where the Hamiltonian of the system was given. The model was generalized to include external dc magnetic \cite{skip15prl,skip18prl} or electric \cite{skip19prb} fields. It has been also used to study photonic crystals that we consider here \cite{antezza13,skip20epj}. Following these previous works, we will study localization properties of quasimodes $\bm{\psi}_m$ of the atomic system found as eigenvectors of a $3N \times 3N$  Green's matrix ${\hat G}$:
\begin{eqnarray}
{\hat G} \bm{\psi}_m = \Lambda_m \bm{\psi}_m,\;\;\;\; m = 1,\ldots,3N.
\label{eigen}
\end{eqnarray}
The matrix ${\hat G}$ describes the coupling between the atoms via the electromagnetic waves (light) and is composed of $N \times N$ blocks of size $3 \times 3$. A block ${\hat G}_{jn}$ gives the electric field created at a position $\vec{r}_n$ of the atom $n$ by an oscillating point dipole at a position $\vec{r}_j$ of the atom $j$ ($j$, $n = 1, \ldots, N$). It has elements
\begin{eqnarray}
G_{jn}^{\mu \nu} &=& i\delta_{jn} \delta_{\mu \nu}
+ (1 - \delta_{jn}) \frac{3}{2}
\frac{e^{i k_0 r_{jn}}}{k_0 r_{jn}}
\nonumber \\
&\times& \left[ P(i k_0 r_{jn}) \delta_{\mu \nu}
+ Q(i k_0 r_{jn})
\frac{r_{jn}^{\mu} r_{jn}^{\nu}}{(r_{jn})^2} \right],
\label{green}
\end{eqnarray}
where $P(x) = 1-1/x+1/x^2$, $Q(x) = -1+3/x-3/x^2$, $\vec{r}_{jn} = \vec{r}_n - \vec{r}_j$, and the indices $\mu$, $\nu = x, y, z$ denote the projections of $\vec{r}_{jn}$ on the axes $x$, $y$, $z$ of the Cartesian coordinate system: $r_{jn}^{x} = x_{jn}$, $r_{jn}^{y} = y_{jn}$, $r_{jn}^{z} = z_{jn}$.
\red{The inverse of the resonant wave number of an isolated atom $k_0 = \omega_0/c$ provides a convenient length scale by which we will normalize all other length scales. Here $c$ is the speed of light in the free space.}

\begin{figure*}[t]
\includegraphics[width=0.9\textwidth]{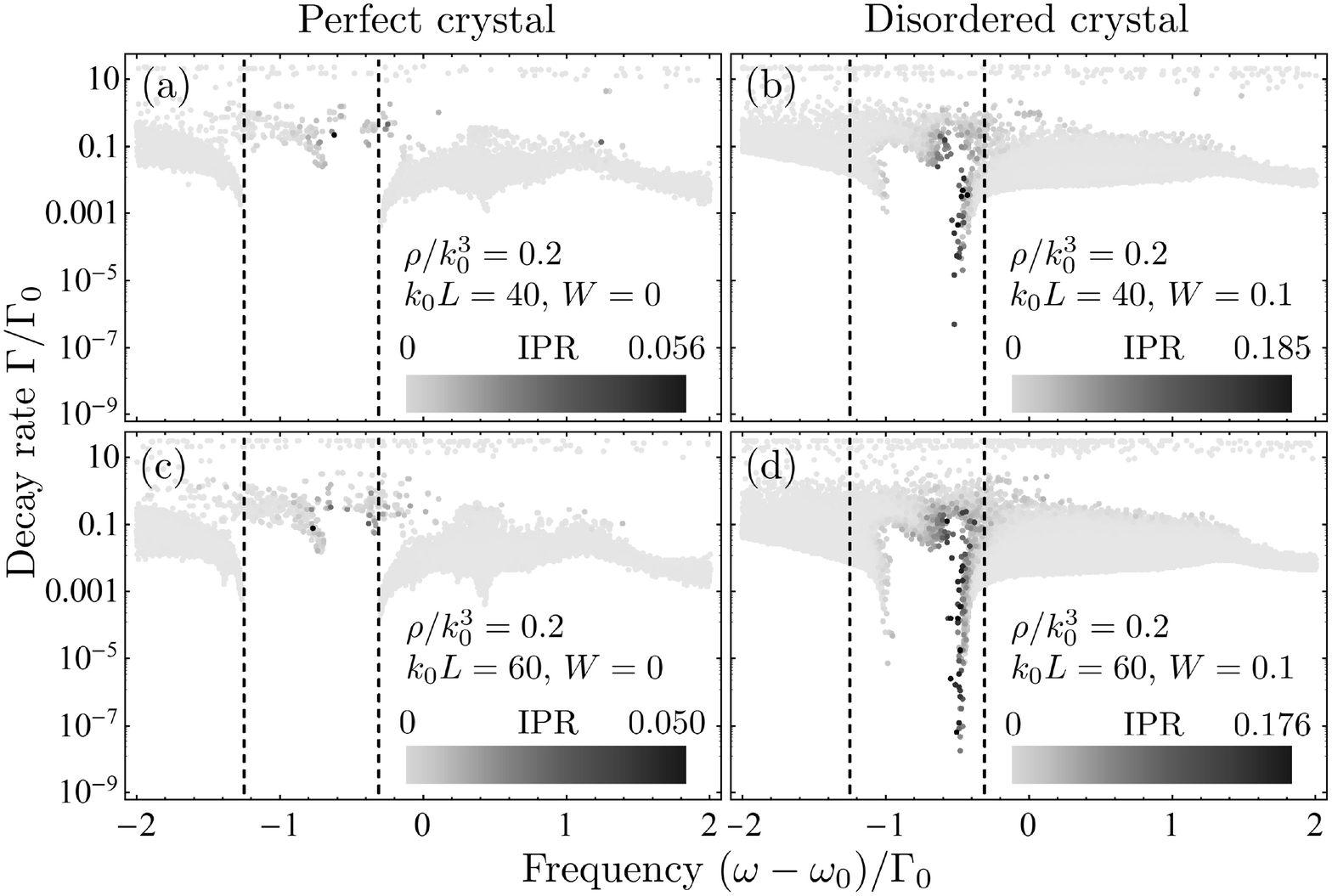}
\vspace{-5mm}
\caption{\label{fig_ipr_perfect}
Eigenvalues of a single realization of the Green's matrix for perfect [left column, panels (a) and (c)] and disordered [right column, panels (b) and (d)] diamond crystals of two different sizes $k_0 L = 40$ (upper row) and 60 (lower row). Each point in the graph corresponds to an eigenvalue and its grey scale represents the IPR of the corresponding eigenvector, from light grey for IPR = 0 (extended eigenvectors) to black for the maximum IPR (different for each panel, most localized eigenvectors). Vertical dashed lines show band edges. Only a part of the eigenvalue spectrum $(\omega-\omega_0)/\Gamma_0 \in [-2,2]$ is shown.
}
\end{figure*}

An eigenvector $\bm{\psi}_m = (\psi_m^{1}, \ldots, \psi_m^{3N})^T$ of the matrix ${\hat G}$ describes the spatial structure of the $m$-th quasimode: $\psi_m^{3(j-1)+\mu}$ gives the $\mu$-th component of the electric field on the atom $j$. The corresponding eigenvalue $\Lambda_m$ yields the eigenfrequency $\omega_m$ and the decay rate $\Gamma_m/2$ of the quasimode:
$\omega_m = \omega_0 - (\Gamma_0/2) \mathrm{Re} \Lambda_m$ and $\Gamma_m/2 = (\Gamma_0/2) \mathrm{Im} \Lambda_m$.
Spatial localization of quasimodes can be quantified by the so-called inverse participation ratio (IPR):
\begin{eqnarray}
\mathrm{IPR}_m = \sum\limits_{j=1}^N \left\{\sum\limits_{\mu=1}^3 \left| \psi_m^{3(j-1)+\mu} \right|^2 \right\}^2,
\label{ipr}
\end{eqnarray}
where we assume that the eigenvectors $\bm{\psi}_m$ are normalized:
\begin{eqnarray}
\sum\limits_{j=1}^N \sum\limits_{\mu=1}^3 \left| \psi_m^{3(j-1)+\mu} \right|^2 = 1.
\label{norm}
\end{eqnarray}
It is easy to see that $\mathrm{IPR}_m = 1$ for a state localized on a single atom and  $\mathrm{IPR}_m = 1/N$ for a state that is uniformly delocalized over all $N$ atoms of the system. Generally, $\mathrm{IPR}_m \sim 1/M$ for a state localized on $M$ atoms.

The spectral distribution of quasimodes can be characterized by the density of states (DOS) ${\cal N}(\omega)$ defined in an open system as \cite{skip20epj,caze13}
\begin{eqnarray}
{\cal N}(\omega) = \frac{1}{3 N \pi}\sum\limits_{m=1}^{3N} \frac{(\Gamma_m/2)}{(\omega - \omega_m)^2 + (\Gamma_m/2)^2}.
\label{dos}
\end{eqnarray}
${\cal N}(\omega)$ is normalized such that the number of states inside an infinitely narrow frequency interval $d\omega$ is
$dN = 3N {\cal N}(\omega) d\omega$.
\red{
Thanks to such a normalization, ${\cal N}(\omega)$ converges to a limiting shape corresponding to the infinite crystal as the size of the crystal increases \cite{skip20epj}. Note that in our formalism, the number of quasimodes is equal to the size $3N$ of the matrix ${\hat G}$ and hence increases with $N$ for all frequencies, including those inside the band gap. However, as discussed elsewhere \cite{skip20epj}, the quasimodes corresponding to the frequencies inside the band gap are confined near the crystal boundary and hence their number grows proportionally to the crystal surface $\pi L^2 \propto N^{2/3}$. This growth is slower than the growth of the total number of modes and hence the relative weight of these quasimodes tends to zero in the thermodynamic limit $N \to \infty$ and ${\cal N}(\omega) \propto 1/L$ \cite{skip20epj,bin18}.}

\red{In this paper, we will present results for crystals of four different sizes $k_0 L = 30$, 40, 50 and 60 composed of $N = 2869$, 6851, 13331 and 22929 atoms, respectively. These numbers of atoms have been adjusted to maintain the same lattice constant $k_0 a = 3.4$ and the same average atomic number density $\rho/k_0^3 = 0.2$.} The lattice constant is chosen small enough for a band gap to open in the spectrum of the ideal lattice \cite{antezza09pra}
\red{as we illustrate by DOS calculations shown in Fig.\ \ref{fig_dos} for}
the perfect ($W = 0$) and disordered \red{crystals of size $k_0 L = 50$}.
For disordered lattices, DOS has been averaged over many independent random atomic configurations using the Monte Carlo method \cite{binder97}. DOS inside the band gap is different from zero due to the finite size of the considered sample \cite{skip20epj,bin18}. We observe that the band gap narrows when disorder in atomic positions is introduced ($W = 0.1$) and closes for strong enough disorder ($W = 0.2$). No signature of a band gap is found for a fully random system in which the atomic positions $\vec{r}_j$ are chosen randomly inside a sphere without any reference to the periodic diamond structure. Therefore, it turns out that our disordered photonic crystal preserves a band gap only for relatively weak disorder $W < 0.2$.

It is worthwhile to note that DOS ${\cal N}(\omega)$ reflects only the atomic component of elementary excitations of the system comprising the atoms and the electromagnetic field. Thus, low ${\cal N}(\omega)$ does not necessarily correspond to a small number of excitations at a given frequency $\omega$ but can simply mean that the atomic subsystem is weakly involved and the excitations look very much like freely propagating photons. This typically happens far from the atomic resonance, for $|\omega - \omega_0| \gg \Gamma_0$, where the coupling of light with atoms is inefficient. The absence of free-field solutions that have no atomic component for frequencies inside the band gap has been demonstrated previously \cite{klugkist06,antezza09pra}. A gap in ${\cal N}(\omega)$ thus corresponds to a gap in the total density of states and a gap in the density of electromagnetic modes ${\cal N}_{\textrm{EM}}(\omega)$ entering the localization criterion (\ref{loccrit}), even though ${\cal N}(\omega) \ne {\cal N}_{\textrm{EM}}(\omega)$.

\red{
In addition to DOS ${\cal N}(\omega)$, another interesting quantity is the {\em local} density of states (LDOS) ${\cal N}(\omega, \vec{r})$. In a photonic crystal of finite size, LDOS exhibits rapid spatial variations within each unit cell of the crystal and slow overall evolution with the distance to the boundaries \cite{leistikow11,mavidis20}. Disorder introduces fluctuations of LDOS and the statistics of the latter may serve as a criterion for Anderson localization \cite{schubert10}. However, calculation of LDOS for our model would require finding eigenvectors $\bm{\psi}_m$ of the matrix ${\hat G}$ which is a much more time-consuming computational task than finding the eigenvalues $\Lambda_m$ that are needed to calculate ${\cal N}(\omega)$ [see Eq.\ (\ref{dos})]. Even though we present some results for $\bm{\psi}_m$  in Figs.\ \ref{fig_ipr_perfect}--\ref{fig_state} below, their statistical analysis including the calculation of the average LDOS $\langle {\cal N}(\omega, \vec{r}) \rangle$ is beyond the scope of this work.
}

\section{Localized states inside the band gap}
\label{sec:states}

It follows from Fig.\ \ref{fig_dos}(b) that some quasimodes cross \red{over} the edges of the band gap when disorder is introduced in the photonic crystal (compare DOS corresponding to $W = 0$ and $W = 0.1$). In order to study the spatial localization properties of these modes, we show quasimode eigenfrequencies $\omega$ and decay rates $\Gamma$ together with their IPR for the perfect diamond crystal and a single realization of the disordered crystal in Fig.\ \ref{fig_ipr_perfect}. For the perfect crystal [left column of Fig.\ \ref{fig_ipr_perfect}, panels (a) and (c)], the vast majority of the modes both inside and outside the band gap are extended and have $\mathrm{IPR} \sim 1/N$. The distribution of quasimodes on the frequency-decay rate plane changes only slightly upon increasing the size of the system from $k_0L = 40$ to $k_0L = 60$ [compare panels (a) and (c) of Fig.\ \ref{fig_ipr_perfect}]. In contrast, the disordered photonic crystal exhibits some localized modes with appreciable IPR near band edges and in particular near the upper band edge, see Fig.\ \ref{fig_ipr_perfect}(b) and (d). These modes have decay rates (life times) that are significantly smaller (longer) than the decay rates (life times) of any modes of the perfect crystal. In addition, the number of such localized modes increases and their decay rates decrease significantly when the disordered crystal gets bigger [compare panels (b) and (d) of Fig.\ \ref{fig_ipr_perfect}]. Such a combination of spatial localization with small decay rates and the scaling with the sample size is typical for disorder-induced quasimode localization \cite{skip14prl,skip16prb}.

\begin{figure}[t]
\includegraphics[width=\columnwidth]{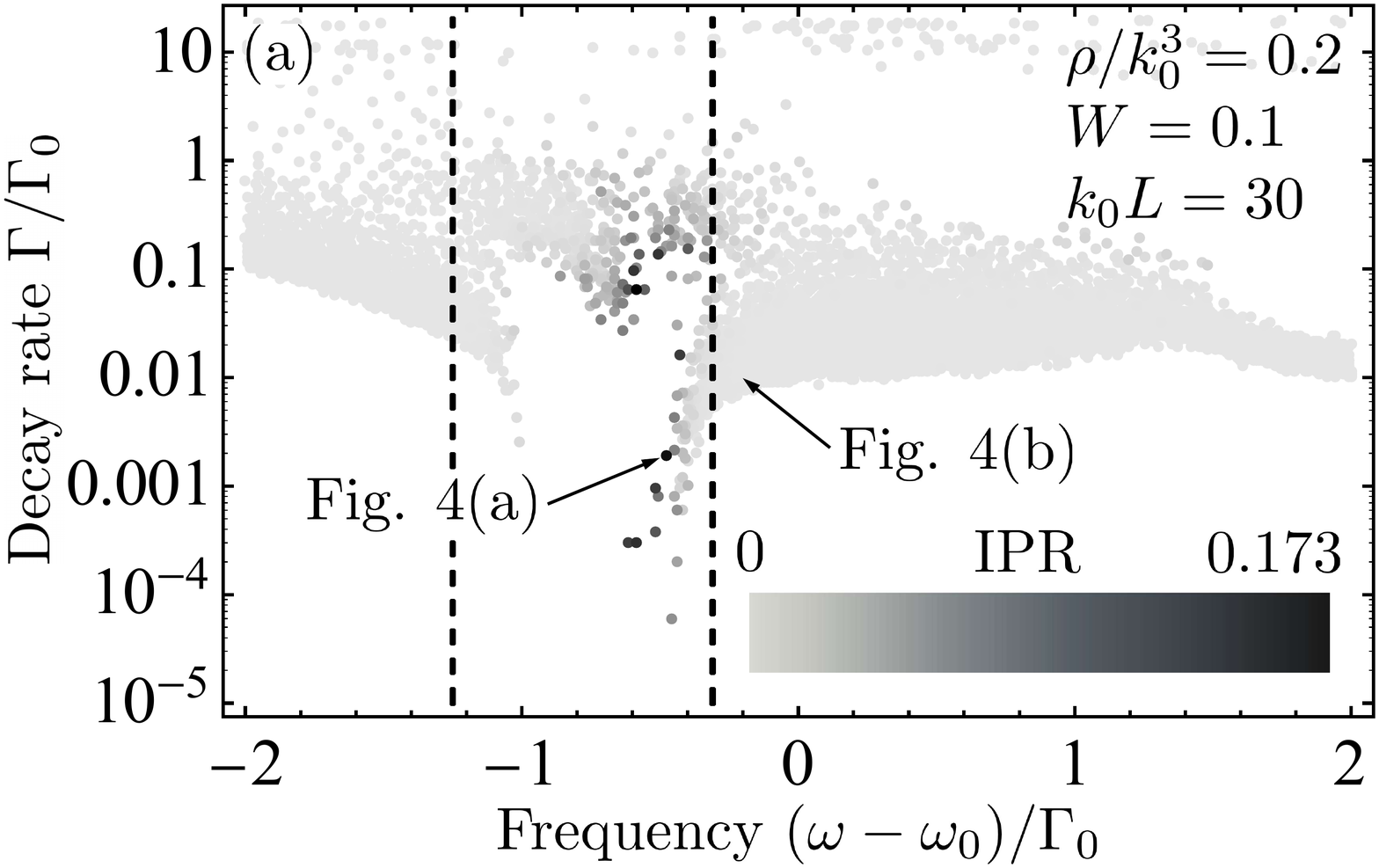}\\
\vspace{-5mm}
\includegraphics[width=\columnwidth]{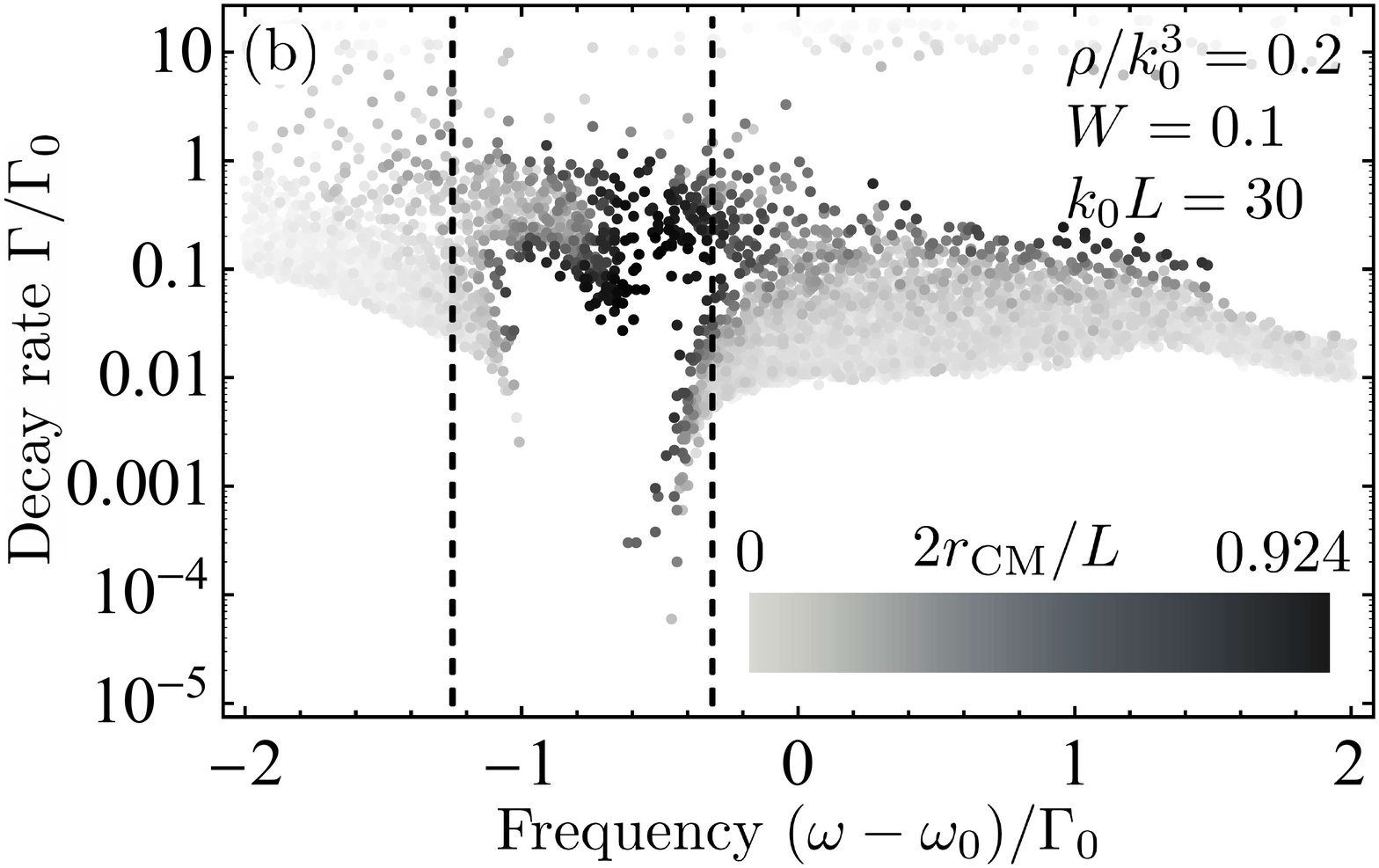}
\vspace{-10mm}
\caption{\label{fig_ipr_cm}
Eigenvalues of a single realization of the Green's matrix for a disordered diamond crystal of size $k_0 L = 30$. Each point in the graph corresponds to an eigenvalue and its grey scale represents the IPR (a) or the center of mass $r_{\mathrm{CM}}$ (b) of the corresponding eigenvector.
\red{The spatial structure of the eigenvectors corresponding to the two eigevalues indicated by arrows in panel (a) is illustrated in Fig.\ \ref{fig_state}.}
Vertical dashed lines show band edges. Only a part of the eigenvalue spectrum $(\omega-\omega_0)/\Gamma_0 \in [-2,2]$ is shown.
}
\end{figure}

\begin{figure}[t]
\includegraphics[width=\columnwidth]{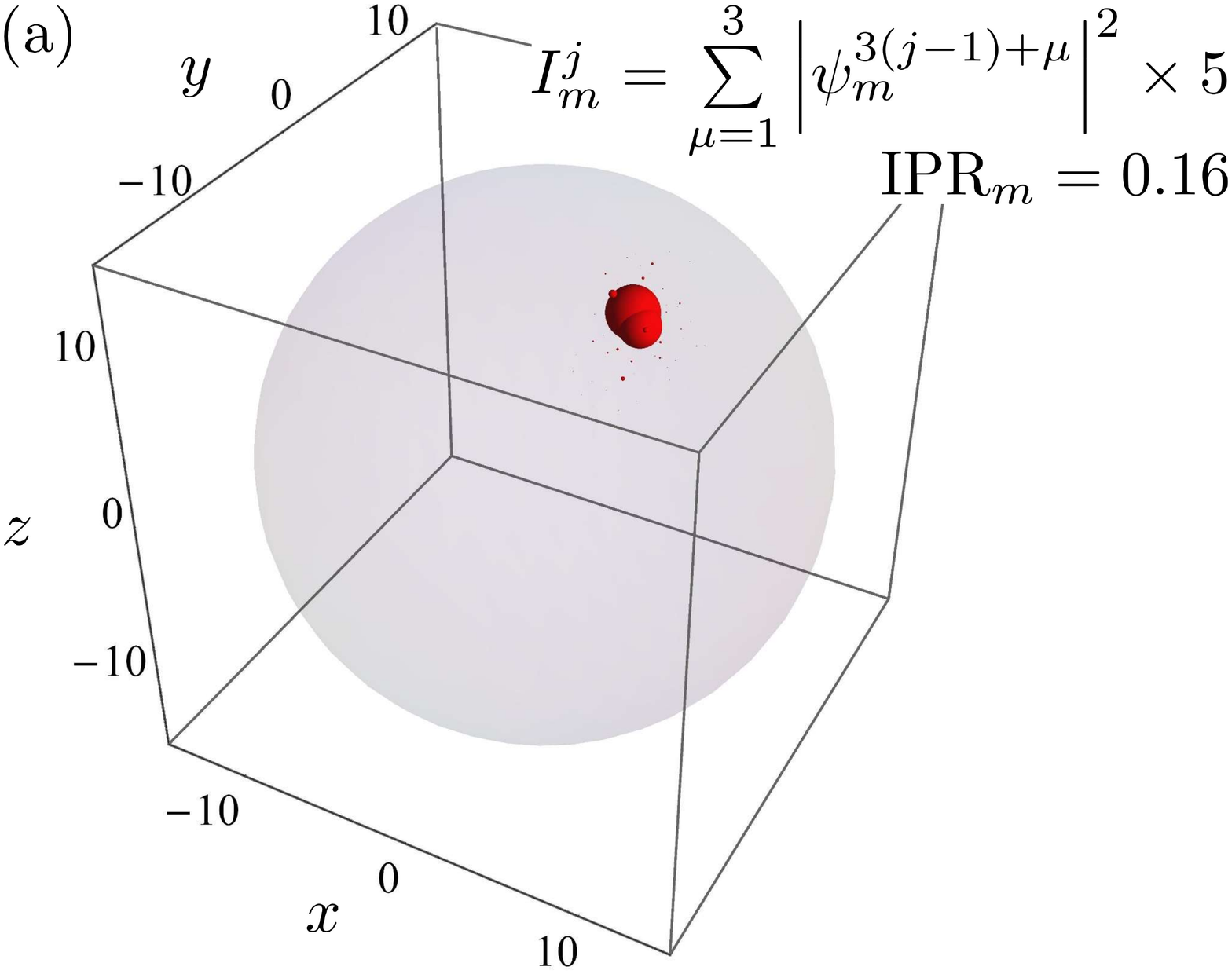}\\
\includegraphics[width=\columnwidth]{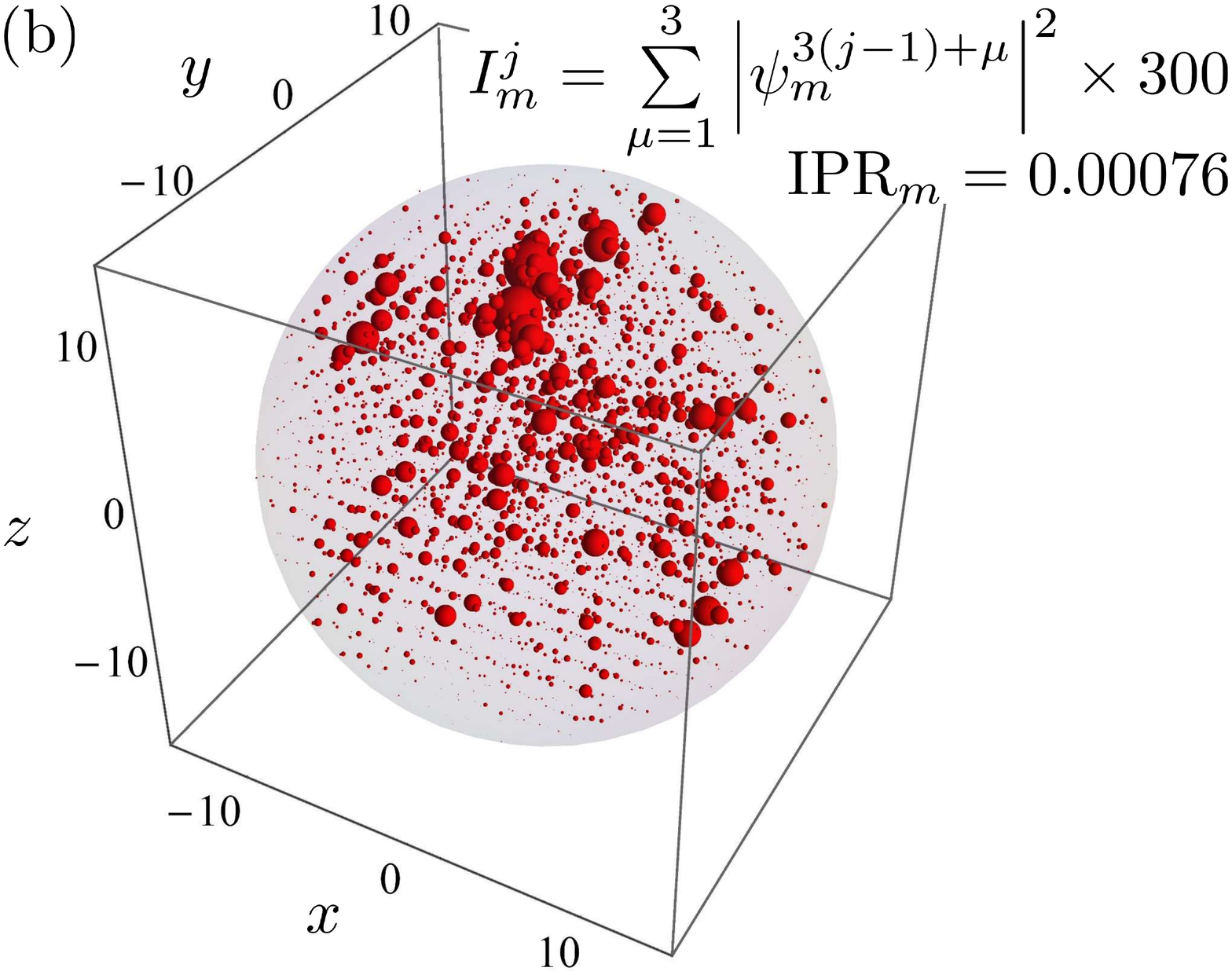}
\vspace{-5mm}
\caption{\label{fig_state}
\red{Visualization of eigenvectors (quasimodes) corresponding to the eigenvalues indicated by arrows in Fig.\ \ref{fig_ipr_cm}(a). A quasimodes $\bm{\psi}_m$ is represented by $N$ red spheres centered at the locations $\vec{r}_j$ ($j = 1, \ldots, N$) of the $N$ atoms and having radii proportional to the intensities
$I_m^j = \sum_{\mu=1}^{3} | \psi_m^{3(j-1)+\mu} |^2$
of the quasimode on the atom. The quasimode (a) is spatially localized and has a relatively high IPR whereas the quasimode (b) is spatially extended. Grey spheres in both panels visualize the spherical region occupied by the disordered photonic crystal.}
}
\end{figure}

In addition to extended modes everywhere in the spectrum, isolated localized modes appear in the middle of the band gap of the perfect crystal [see Fig.\ \ref{fig_ipr_perfect}(a) and (c)]. Their $\mathrm{IPR} \sim 5 \times 10^{-2}$ is small but still considerably larger than $1/N \sim 10^{-4}$ expected for extended modes. Such modes do not disappear and become even more numerous in the disordered crystal [see Fig.\ \ref{fig_ipr_perfect}(b) and (d)]. They differ from the modes near band edges by their much larger decay rates that are virtually independent of the crystal size. Our previous work suggests that all modes in the middle of the band gap of a photonic crystal are confined near the crystal boundary, which may explain their $\mathrm{IPR} \propto 1/N^{2/3} \gg 1/N$ \cite{skip20epj}. In the presence of disorder, some of these modes may, in addition, be restricted to a small part of sample surface \cite{maximo19}, which may explain their larger IPR. To confirm this explanation, we compute the center of mass of a mode $\bm{\psi}_m$ as
\begin{eqnarray}
\vec{r}_{\mathrm{CM}}^{(m)} = \sum\limits_{j=1}^{N} \vec{r}_j \left[ \sum\limits_{\mu=1}^{3} |\psi_m^{3(j-1)+\mu}|^2 \right].
\label{rcm}
\end{eqnarray}
Figure \ref{fig_ipr_cm} shows that the modes in the middle of the band gap, including those having large IPR, tend to have the absolute value of their center of mass $r_{\mathrm{CM}}$ to be of order of the radius $L/2$ of the atomic sample. These modes are therefore confined at the sample boundary as we have anticipated. The confinement at the boundary explains the relatively large decay rates of these modes and the weak dependence of decay rates on the sample size. Although the role of surface modes discussed above may appear to be important in the calculations presented in this work, this is due to the relatively small sizes $k_0 L = 30$--60 of considered atomic samples limited by the computational constraints to which our numerical calculations are subjected. In the limit of $k_0 L \to \infty$ relevant for the analysis of modes localized by disorder in the bulk, surface modes play no role. In finite samples accessible to numerical calculations, the impact of surface modes can be minimized by using a scaling analysis presented in the next section.
\red{The need for a scaling analysis is also due to the absence of a univocal relation between the decay rate $\Gamma$ of a quasimode and its localization properties. Indeed, some of the black points in Fig.\ \ref{fig_ipr_cm}(a) correspond to much larger $\Gamma$ than some of the grey points, showing that the IPR and $\Gamma$ are not directly related. However, a relation can be established between the scaling of (normalized) $\Gamma$ with the sample size $L$ and the spatial localization of quasimodes at a given frequency.} Surface states do not follow the same scaling with the sample size as the modes localized in the bulk, which provides a way of discriminating between these two types of modes.

\red{
Similarly to the panels (b) and (d) of Fig.\ \ref{fig_ipr_perfect}, Fig.\ \ref{fig_ipr_cm}(a) shows that quasimodes with large IPR appear inside the band gap of the photonic crystal due to disorder. The spatial structure of these spatially localized quasimodes is very different from that of the extended quasimodes with frequencies outside the band gap, as we illustrate in Fig.\ \ref{fig_state}.
}

\section{Finite-size scaling}
\label{sec:scaling}

The finite-size scaling analysis is a way to access the behavior of an infinitely large system from the experimental or numerical data available for finite-size systems only. It is a common approach for analysis of phase transitions \cite{privman90,binder97} and has been widely used to characterize Anderson localization transitions in electronic \cite{mackinnon81,shklovskii93,slevin01,slevin14}, optical \cite{sheikhan09,skip18prl} and mechanical \cite{pinski12epl,pinski12jpcm} systems. Very generally, one chooses a quantity (let's denote it by $\Omega$) that is supposed to take two very different values (say, 0 and $\infty$) for the infinitely large system in the two different phases. The behavior of the quantity $\Omega$ is then studied as a function of sample size $L$ and regions of the parameter space are identified in which $\Omega$ increases or decreases with $L$. A point (for 1D parameter space), a line (2D), or a surface (3D) separating these regions is identified as a boundary between the two phases at which $\Omega$ is independent of $L$. Moreover, it often turns out that even when the parameter space of the physical system under consideration is multidimensional, all the parameters can be combined into a single one that is the only relevant near the phase transition point. In this situation known as the `single-parameter scaling' \cite{mackinnon81}, the critical exponents of the transition can be estimated from the behavior of $\Omega$ with $L$ for finite $L$.

In the context of Anderson localization, the (dimensionless) electrical conductance $g$ of a sample of size $L$ was identified as the \red{most} relevant quantity to consider: $\Omega = g$ \cite{abrahams79}. Obviously, the conductance of a 3D metallic cube of side $L$ in which all the electronic eigenstates are extended, $g \propto L$, grows with $L$ whereas one expects a decreasing conductance $g \propto \exp(-L/\xi)$ if the electronic eigenstates are localized at the scale of localization length $\xi$ and the sample is an (Anderson) insulator. We thus see that in the limit of $L \to \infty$, $g \to \infty$ if the eigenstates are extended and $g \to 0$ if they are spatially localized. In addition, one expects $g$ to be independent of $L$ at the critical point \cite{abrahams79}. This is, by the way, the essence of the Thouless criterion of Anderson localization $g \sim \mathrm{const}$ \cite{thouless77,abrahams79}, where `const' is a number of order unity.

The conceptual picture described above needs some adjustments when it comes to its application to the physical reality. Indeed, in a disordered system, $g$ is a random quantity and it is not clear how exactly its scaling with the sample size should be understood \cite{shapiro86}. The simplest option of analyzing its average value $\langle g \rangle$ turned out to be not always appropriate because $\langle g \rangle$ may be dominated by rare realizations of disorder with large $g$ \cite{shapiro86,cohen88}. Another, more intelligent guess is to use the average of the logarithm of $g$, $\langle \ln g \rangle$. This indeed allows to obtain reasonable results \cite{slevin01} but has the weakness of being somewhat arbitrary as a choice: why $\langle \ln g \rangle$ and not $\langle (\ln g)^2 \rangle$, $\langle (\ln g)^3 \rangle$ or the mean value of some other function of $g$? Although averaging different functions of $g$ may yield identical results for the critical properties of the localization transition in some models \cite{slevin01}, it is not so for the model of point scatterers considered here \cite{skip16prb}. This is why studying the full probability distribution function $P(g)$ instead of statistical moments of $g$ or $\ln g$ is necessary \cite{shapiro86,cohen88}.
\red{Conductance $g$ and its probability distribution function $P(g)$ are not the only quantities that can be used for the scaling analysis of the Anderson transition. Alternatives include the distribution of eigenvalue (level) spacings \cite{shklovskii93} or the multifractal spectrum \cite{rodriguez10} as the most prominent examples. Note that although initially proposed for Hermitian systems \cite{shklovskii93}, the finite-size scaling of spacings between eigenvalues has been recently extended to the non-Hermitian case \cite{tzor20,huang20} and thus can, in principle, be applied to analyze open disordered systems as the one considered in this work. However, $g$ and $P(g)$ still remain the most simple and computationally accessible quantities to analyze.}

Conductance as a ratio of the electric current to the voltage that causes it, is a notion that is proper to electronics and seems to be impossible to generalize to light. However, Thouless has noticed that if one divides the typical decay rate $\Gamma/2$ of quasimodes of an open quantum or wave system by the average spacing between quasimode frequencies $\Delta \omega$, the resulting `Thouless conductance' is equal to the electrical conductance $g$ for a metal wire: $(\Gamma/2)/\Delta\omega = g$ \cite{thouless77}. The advantage of the Thouless definition is that it can be readily generalized to any waves independent of any electrical currents or potential differences in the considered physical system. In our open, finite-size photonic crystal we define
\begin{eqnarray}
g_m = \frac{\Gamma_m/2}{\langle |\omega_m - \omega_{m-1}| \rangle}
= \frac{\mathrm{Im}\Lambda_m}{\langle |\mathrm{Re}\Lambda_m - \mathrm{Re}\Lambda_{m-1}| \rangle},
\label{cond}
\end{eqnarray}
where the eigenfrequencies $\omega_m$ are assumed to be ordered. We note that in a closed system the matrix ${\hat{G}}$ would be Hermitian and its eigenvalues real. Then the denominator of Eq.\ (\ref{cond}) would be equal to $1/[3N {\cal N}(\omega)]$. However, in the open system that we consider, the relation between the average spacing between eigenfrequencies $\omega_m$ and DOS is only approximate because the definition of DOS (\ref{dos}) involves decay rates of quasimodes as well. In practice, we can still approximately write
\begin{eqnarray}
g_m \simeq \frac{\Gamma_m}{2} {\cal N}(\omega_m) 3N.
\label{cond2}
\end{eqnarray}
Using this definition instead of Eq.\ (\ref{cond}) would barely modify the results following from the finite-size scaling analysis below because neither $\langle |\omega_m - \omega_{m-1}| \rangle^{-1}$ nor ${\cal N}(\omega)$ exhibit singularities at the localization transition points.

\begin{figure}[t]
\includegraphics[width=\columnwidth]{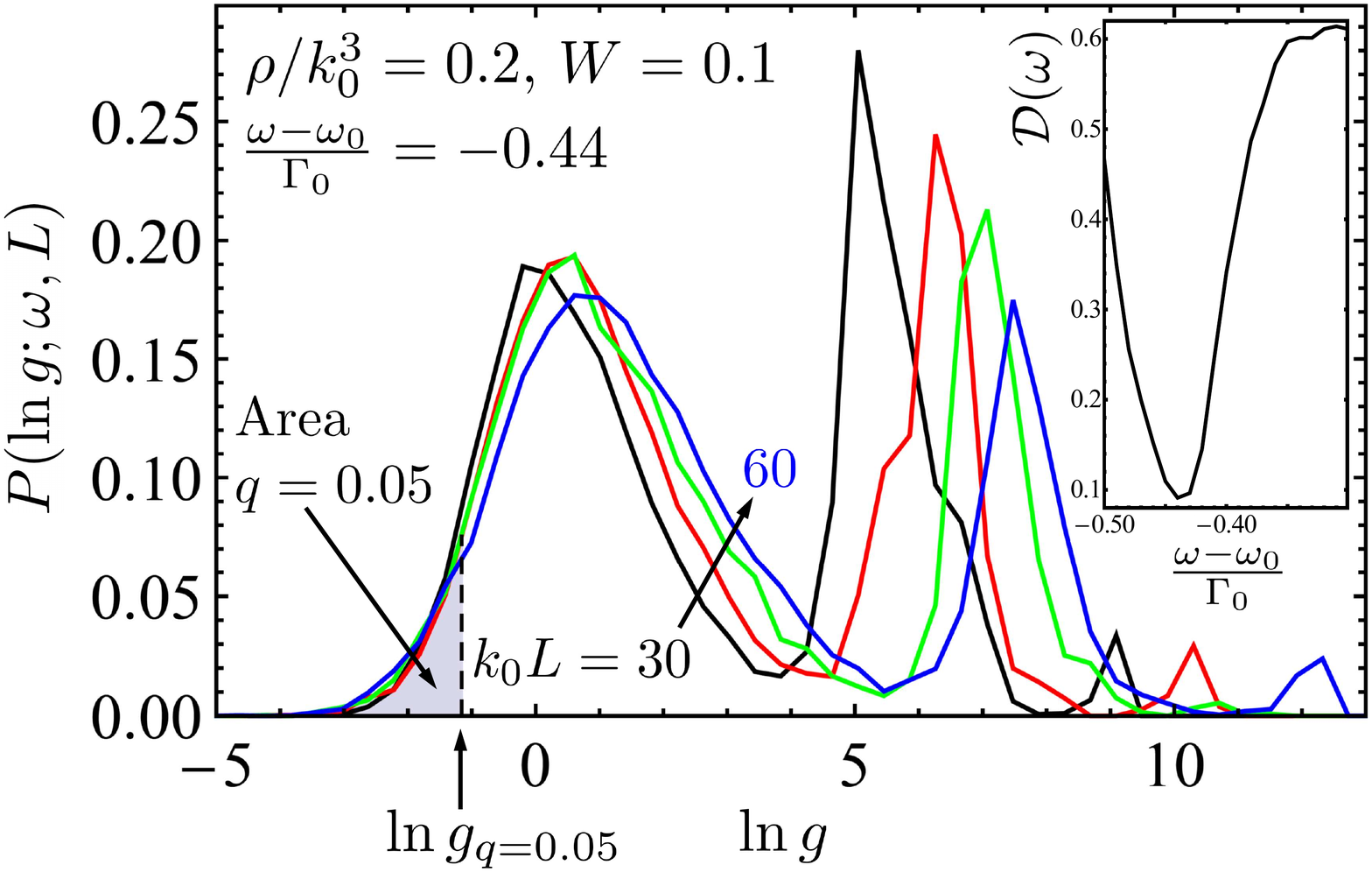}\\
\vspace{-7mm}
\caption{\label{fig_distr}
Probability density of the logarithm of the Thouless conductance $g$ at the critical point of the localization transition for different sizes of the disordered crystal: $k_0 L = 30$ (black), 40 (red), 50 (green), 60 (blue). The numbers of random realizations of the matrix ${\hat G}$ used for different sizes are \red{2200}, \red{900}, \red{461} and \red{180}, respectively. All eigenvalues within a frequency interval of width $0.01 \Gamma_0$ around $\omega - \omega_0 = -0.44 \Gamma_0$ are used to estimate $P(\ln g; \omega, L)$. Probability densities corresponding to different sizes coincide for small $g$; the grey shaded area below $P(\ln g; \omega, L)$ illustrates the notion of $q$-th percentile $\ln g_q$ for the firth percentile $q = 0.05$. Inset: the distance ${\cal D}(\omega)$ between probability densities corresponding to $k_0 L = 30$ and 60 attains a minimum at the critical point $(\omega-\omega_0)/\Gamma_0 \simeq -0.44$.
\red{The step of frequency discretization is $0.01 \Gamma_0$ for this figure.}}
\end{figure}

In a disordered photonic crystal, the Thouless conductance defined by Eq.\ (\ref{cond}) is a random quantity and at fixed scatterer density $\rho$ and disorder strength $W$, its statistical properties can be characterized by a probability density function $P(\ln g; \omega, L)$. Here we choose to work with $\ln g$ instead of $g$ because $g$ varies in a rather wide range. The probability density is parameterized by the frequency $\omega$ of the quasimodes and the sample size $L$. We estimate $P(\ln g; \omega, L)$ for different $\omega$ around the upper edge of the band gap observed in Fig.\ \ref{fig_dos} by numerically diagonalizing many independent random realizations of the matrix ${\hat G}$ for different sizes $L$ of the disordered photonic crystal. Figure \ref{fig_distr} shows the results for $W = 0.1$ and a particular frequency $\omega = \omega_0 - 0.44 \Gamma_0$ for which the so-called Harald Cram\'{e}r's distance between probability density functions corresponding to the smallest and largest $L$ is minimized (see the inset of Fig.\ \ref{fig_distr}). The Harald Cram\'{e}r's distance is
\begin{eqnarray}
{\cal D}(\omega) &=& \int\limits_{-\infty}^{\infty} d(\ln g)
\left| P(\ln g; \omega, L = 30 k_0^{-1})
\right. \nonumber \\
&-& \left.  P(\ln g; \omega, L = 60 k_0^{-1}) \right|^2.
\label{dif}
\end{eqnarray}

Interestingly, the frequency $\omega$ for which ${\cal D}(\omega)$ is minimal also corresponds to the frequency for which distributions $P(\ln g; \omega, L)$ corresponding to different $L$ tend to coincide for small $g$, see the main panel of Fig.\ \ref{fig_distr}. Following our previous work \cite{skip16prb}, we identify this relative $L$-independence of $P(\ln g; \omega, L)$ as a signature of a critical point of a localization transition (also called a mobility edge).
\red{The probability density of conductance near the transition from extended to localized states has been extensively studied in the past for both quasi-1D \cite{muttalib99,froufe02} and 3D \cite{markos93,markos99,muttalib05} disordered systems without band gaps. For small $g$, our $P(\ln g; \omega, L)$ exhibits a tail decreasing to zero as $g \to 0$ in agreements with the previous prediction \cite{markos99}. However, in contrast to the expectations \cite{markos93,markos99,muttalib05}, our  $P(\ln g; \omega, L)$ does not have a smooth, size-independent shape for large $g$.} We attribute this fact to the following reason. The realistic physical model of two-level atoms arranged in a diamond lattice that we consider, may exhibit other physical phenomena in addition to the eigenmode localization near band edges. These phenomena may be due, for example, to the collective interaction between atoms (sub- \cite{guerin16,weiss18,moreira19} and superrradiance \cite{araujo16,cottier18}) or to the specific structure of their spatial arrangement (potential topological phenomena \cite{ryu09,takahashi17}). Without having any relation to quasimode localization, these phenomena may cause some particular features of $P(\ln g; \omega, L)$ and exhibit some $L$-dependence. Some of these features may disappear in the limit of $k_0 L \to \infty$ but it is impossible to claim such a disappearance from our calculations performed for finite $k_0 L = 30$--60, which are likely to be insufficient to clearly observe the behavior expected in the limit of $k_0 L \to \infty$. For example, we see from Fig.\ \ref{fig_distr} that $P(\ln g; \omega, L)$ exhibits a pronounced peak at $\ln g \gtrsim 5$. \red{The peak} shifts to larger $g$ and reduces in magnitude as $L$ increases. This peak corresponds to superradiant states with short lifetimes which always exist in a finite-size system but which have a statistical weight decreasing with $L$. It is likely that the peak would vanish in the limit of $L \to \infty$ which is, however, inaccessible for our calculations.

We will use the small-$g$ part of $P(\ln g; \omega, L)$ that becomes $L$-independent at $\omega \simeq \omega_0 - 0.44 \Gamma_0$ (see Fig.\ \ref{fig_distr}), to quantify the localization transition. The finite-size scaling analysis of $P(\ln g; \omega, L)$ can be conveniently performed by analyzing its percentiles $\ln g_q$ \cite{slevin03}. The $q$-th percentile $\ln g_q$ is defined by a relation:
\begin{eqnarray}
q = \int\limits_{-\infty}^{\ln g_q} P(\ln g; \omega, L) d(\ln g)
\label{perc}
\end{eqnarray}
illustrated in Fig.\ \ref{fig_distr} for $q = 0.05$ (firth percentile). Independence of the small-$g$ part of $P(\ln g; \omega, L)$ of $L$ implies that $\ln g_q$ should be $L$-independent for small $q$ as well. Visual inspection of Fig.\ \ref{fig_distr} suggests that $q = 0.05$ is more or less the maximal value of $q$ for which the $L$-independence of $P(\ln g; \omega, L)$ can be assumed. For larger $q$, the dashed vertical line in Fig.\ \ref{fig_distr} would shift to the right and enter into the range of $\ln g$ in which $P(\ln g; \omega, L)$ corresponding to different $L$ are clearly different. The grey shaded area $q$ on the left from the dashed line would then be ill-defined.

\begin{figure*}[t]
\includegraphics[width=0.9\textwidth]{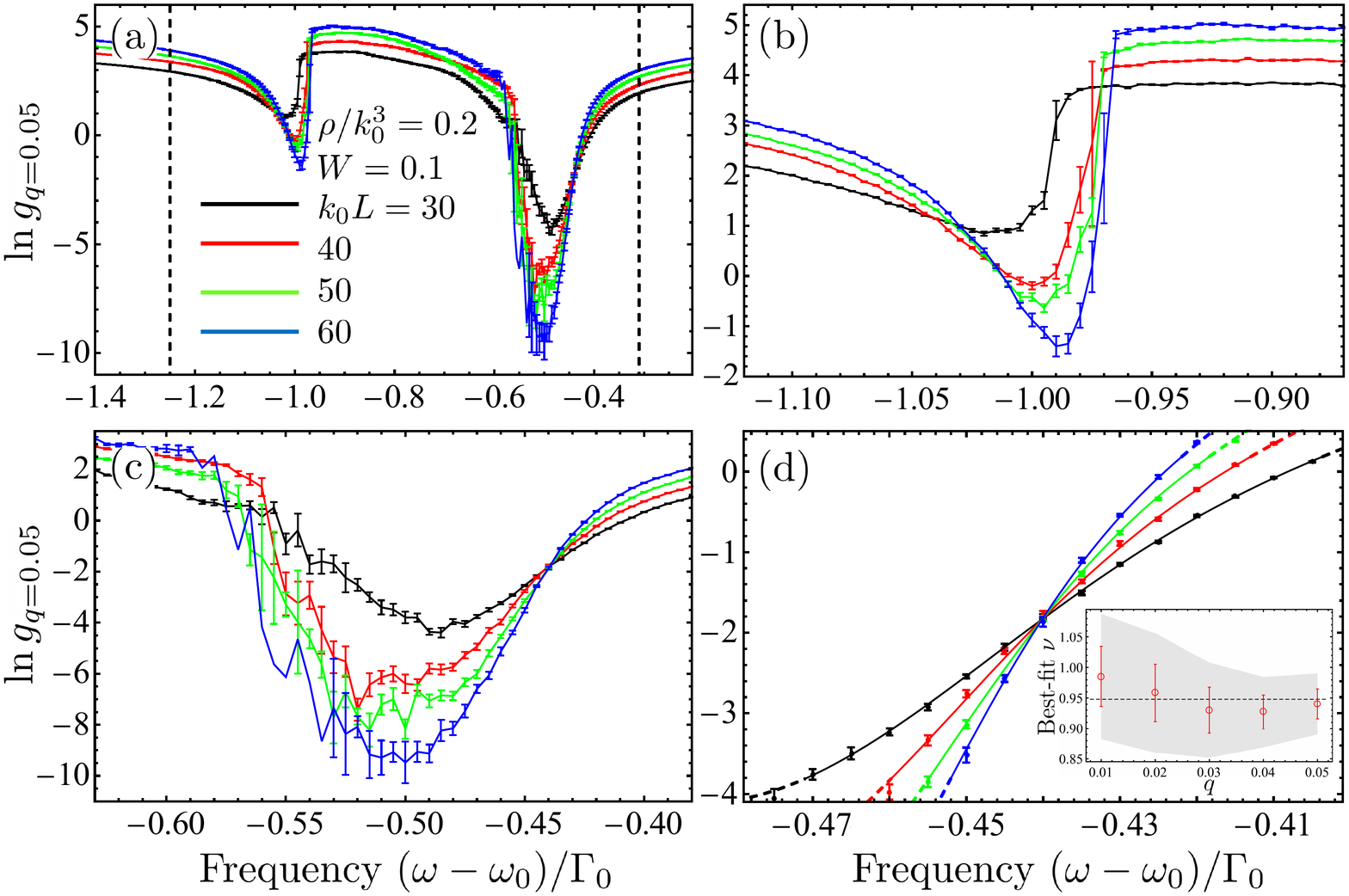}
\vspace{-5mm}
\caption{\label{fig_scaling}
(a) Firth percentile $\ln g_{q = 0.05}$ of the Thouless conductance as a function of frequency $\omega$ for four different sizes $k_0 L$ of the disordered photonic crystal. Very large error bars in the range $(\omega-\omega_0)/\Gamma_0 \in (-0.58, -0.54)$ are not shown. Vertical dashed lines show the band edges. Panels (b) and (c) zoom on the spectral ranges in which $\ln g_{q = 0.05}$ drops near the lower and upper band edges, respectively. (d) Finite-size scaling analysis of the localization transition taking place at $\omega = \omega_c \simeq \omega_0 - 0.44\Gamma_0$ where curves corresponding to different crystal sizes cross in a single point $\{ (\omega_c-\omega_0)/\Gamma_0, \ln g_q^{(c)} \}$. Solid lines represent a joint polynomial fit of Eq.\ (\ref{fit3}) with $m = n = 3$ to the numerical data, dashed lines show their extrapolation beyond the range of data $\ln g_q \in [\ln g_q^{(c)} - \delta(\ln g_q), \ln g_q^{(c)} + \delta(\ln g_q)]$ used for the fit. $\delta(\ln g_q) = 2$ for this figure. The inset shows the best-fit values of the critical exponent $\nu$ for $q = 0.01$--$0.05$ with errors bars corresponding to the standard deviation, the grey area showing the 95\% confidence interval, and the dashed horizontal line indicating the average of $\nu$ over $q$.}
\end{figure*}

We have computed and analyzed \red{the} percentiles $\ln g_q$ for $q = 0.01$--0.05 and present the results for $q = 0.05$ in Fig.\ \ref{fig_scaling}. The results for smaller $q$ are similar but exhibit stronger fluctuations and larger error bars due to poorer statistics. As discussed above, crossings between $\ln g_q$ corresponding to different $L$ are potential signatures of localization transitions. Figure \ref{fig_scaling}(a) suggests that there are two pairs of such crossings, a pair near the lower edge of the band gap and another pair near the upper edge. Panels (b) and (c) zoom on the corresponding frequency ranges. Let us discuss the behavior with increasing the frequency $\omega$. First, a transition to localized states can be identified around $(\omega - \omega_0)/\Gamma_0 \simeq -1.015$ where a common crossing of $\ln g_q$ corresponding to $k_0 L = 40$, 50 and 60 takes place. The line corresponding to $k_0 L = 30$ does not pass through this common crossing point, most probably because this sample size is insufficient to observe the expected large-sample behavior. $\ln g_q$ remains a decreasing function of $L$ for $(\omega - \omega_0)/\Gamma_0 \gtrsim -1.015$ and up to $(\omega - \omega_0)/\Gamma_0 \simeq -0.97$. This is consistent with the appearance of states localized in the bulk of the disordered crystal at frequencies near a band egde (see Figs.\ \ref{fig_ipr_perfect} and \ref{fig_ipr_cm}). The states with frequencies in the middle of the band gap,  $-0.97 \lesssim (\omega - \omega_0)/\Gamma_0 \lesssim -0.57$ in Fig.\ \ref{fig_scaling}(a), appear as relatively localized according to their IPR in Figs.\ \ref{fig_ipr_perfect} and \ref{fig_ipr_cm} but show a scaling behavior that identify them as extended (i.e., $\ln g_q$ grows with $L$). This is consistent with their surface nature: indeed, surface states are restricted to the boundary of the sample and hence the number of atoms on which they have significant amplitudes grows as $L^2$ instead of $L^3$ for extended states in the bulk. Thus, they have larger IPR as compared to the extended states in the bulk, but this IPR still decreases with $L$ (as IPR $\propto 1/L^2$ instead of $1/L^3$). This decrease is reflected in the growth of $\ln g_q$ shown in Fig.\ \ref{fig_scaling}(a). A second band of localized states arises near the upper edge of the band gap, for $-0.57 \lesssim (\omega - \omega_0)/\Gamma_0 \lesssim -0.44$.

Our results for $\ln g_q(\omega, L)$ around $(\omega - \omega_0)/\Gamma_0 \simeq -0.44$ are smooth and have sufficiently small error bars to allow for a quantitative analysis of the transition from localized to extended states. We apply the finite-size scaling procedure to analyze small-$q$ percentiles of $g$ in the framework of the single-parameter scaling hypothesis \cite{slevin03}. The latter postulates that in the vicinity of the localization transition point, $\ln g_q$ is a function of a single parameter $L/\xi(\omega)$, where $|\xi(\omega)|$ is the localization length on the one side from the mobility edge $\omega_c$ and the correlation length on the other side: $\ln g_q(\omega, L) = F_q[L/\xi(\omega)]$. Assuming that the divergence of $\xi(\omega)$ at the transition is power-law, we represent $\xi(\omega)$ as
\begin{eqnarray}
\xi(\omega) = \left[ \sum\limits_{j=1}^{m} A_j w^j \right]^{-\nu}
\label{fit0}
\end{eqnarray}
near $w = (\omega - \omega_c)/\omega_c = 0$. Here $A_j$ are constants and $\nu$ is the critical (localization length) exponent.
We thus can write
\begin{eqnarray}
\ln g_q(\omega, L) = F_q[L/\xi(\omega)] = {\cal F}_q[\psi(\omega, L)]
\label{fit1}
\end{eqnarray}
with a scaling variable
\begin{eqnarray}
\psi(\omega, L) = \left[ \frac{L}{\xi(\omega)} \right]^{1/\nu} = L^{1/\nu} \sum\limits_{j=1}^{m} A_j w^j.
\label{fit2}
\end{eqnarray}
Finally, the scaling function ${\cal F}_q(\psi)$ is expanded in Taylor series:
\begin{eqnarray}
{\cal F}_q(\psi) =  \ln g_q^{(c)} + \sum\limits_{j=1}^{n} B_j \psi^j,
\label{fit3}
\end{eqnarray}
where $\ln g_q^{(c)}$ is the critical value of $\ln g_q$ independent of $L$.

Fits of Eq.\ (\ref{fit3}) to the numerical data are performed with with $\omega_c$, $\ln g_q^{(c)}$, $\nu$, $A_j$ ($j = 1, \ldots, m$), and $B_j$ ($j = 1, \ldots, n$) as free fit parameters. The orders $m$ and $n$ of the expansions (\ref{fit2}) and (\ref{fit3}) are chosen large enough to ensure that the $\chi^2$ statistics
\begin{eqnarray}
\chi^2 = \frac{1}{N_{\mathrm{data}}} \sum\limits_{j=1}^{N_{\mathrm{data}}}
\frac{\{ {\cal F}_q[\psi(\omega_j, L)] - \ln g_q^{(j)} \}^2}{\sigma_j^2}
\label{chi2}
\end{eqnarray}
is of the order 1. Here $N_{\mathrm{data}}$ is the number of data points $\{ \omega_j, \ln g_q^{(j)} \}$ and $\sigma_j$ are statistical errors of $\ln g_q^{(j)}$ shown by error bars in Fig.\ \ref{fig_scaling}. We only fit the numerical data in the range $\ln g_q \in [\ln g_q^{(c)} - \delta(\ln g_q), \ln g_q^{(c)} + \delta(\ln g_q)]$ around the critical value $\ln g_q^{(c)}$ estimated in advance by looking for the minimum of the sum of squares of differences between points corresponding to different $L$.

A joint fit to the numerical data corresponding to four different values of $L$ \red{and $q = 0.05$} is shown in Fig.\ \ref{fig_scaling}(d). It yields $\omega_c = \red{-0.4401 \pm 0.0003}$ and $\nu = \red{0.94 \pm 0.02}$ as the best fit parameters. We repeated the fits for other values of $q$ in the range from 0.01 to 0.05 with the same frequency resolution $0.005 \Gamma_0$ as in Fig.\ \ref{fig_scaling}(d) [see the inset of Fig.\ \ref{fig_scaling}(d) for the best-fit $\nu$] and with a twice finer resolution
\red{and $\delta(\ln g_q) = 1$ instead of $\delta(\ln g_q) = 2$ in Fig.\ \ref{fig_scaling}(d). In addition, we varied the orders $m$ and $n$ of the series expansions (\ref{fit2}) and (\ref{fit3}) from 1 to 3 and introduced an additional, irrelevant scaling variable \cite{slevin14}.} All fits yield consistent values of $(\omega_c - \omega_0)/\Gamma_0$ in the range \red{[$-0.441$, $-0.436$]}. The best-\red{fit} values of the critical exponent are more scattered but remain in the range $\nu = 0.8$--1.1, with large uncertainties up to 20\% for the narrower data range $\delta(\ln g_q) = 1$.

\section{Discussion}
\label{sec:disc}

Whereas the position of the mobility edge found from the finite-size scaling analysis agrees well with the estimation following from the analysis of $P(\ln g; \omega, L)$ (see Fig.\ \ref{fig_distr}), the value of the critical exponent $\nu$ turns out to be well below $\nu_{\text{AM}} \simeq 1.57$ found numerically for the Anderson model (AM) in the 3D orthogonal symmetry class and believed to be universal for disorder-induced localization transitions in 3D systems in the absence of any particular symmetry breaking mechanisms \cite{slevin14}. Cold-atom experiments mimicking the so-called quasiperiodic kicked rotor model indeed yielded values of $\nu$ compatible with $\nu_{\text{AM}}$ \cite{chabe08}, but $\nu \lesssim 1$ significantly different from $\nu_{\text{AM}}$ were reported in low-temperature electron transport experiments in doped semiconductors \cite{thomas85,itoh04}. Recently, values of $\nu \lesssim 1$ have been also found in numerical simulations and attributed to the differences between the physics of real materials and that of the paradigmatic Anderson model and, in particular, to the hybridization of conduction and impurity bands \cite{carnio19}. In our optical problem, the impurity band (i.e., the modes appearing in the band gap due to disorder $W \ne 0$)  is not clearly separated from the band of propagating modes (i.e., the modes in the bands of the perfect crystal) either (see Fig.\ \ref{fig_dos}). This may be a reason for the value of the critical exponent $\nu$ different from $\nu_{\text{AM}}$. Other possible reasons may include a strong anisotropy of optical properties of a photonic crystal near a band edge due to the fact that the first modes that become allowed upon crossing a band edge propagate only in certain directions, and, of course, the vector nature of light of which the full impact on Anderson localization still remains to be understood.

To determine the precise value of $\nu$ and to obtain a better estimation of its uncertainty, more accurate calculations are required. Unfortunately, such calculations are difficult to perform using our approach. Indeed, the approach is based on the diagonalization of large $3N \times 3N$ non-Hermitian matrices ${\hat G}$ and has an advantage of yielding the whole spectrum of a single realization of an open disordered system at once. The downsides of this are that (i) the approach does not allow for focusing on a particular frequency range at a lower computational cost and (ii) studying large systems ($N \gtrsim 10^4$) is computationally expensive. Because the localization transition takes place in a narrow frequency range, only a small fraction of eigenvalues obtained by the numerical diagonalization of ${\hat G}$ is actually used for the estimation of $\nu$.
\red{Indeed, in Fig.\ \ref{fig_scaling}(d) we have chosen to analyze the behavior of $\ln g_q$ within an interval $\ln g_q^{(c)} \pm 2$, which restricts the number of eigenvalues of ${\hat G}$ used in the calculations of $\omega_c$ and $\nu$ to less than 1\%\ of the total number of eigenvalues. Narrowing the interval of considered $\ln g_q$ only decreases the fraction of useful eigenvalues whereas expanding this interval and using more eigenvalues would correspond to leaving the critical regime and hence is not desirable.} Thus, significantly increasing the statistical accuracy of \red{calculations} requires large amounts of computations. Although this drawback of our approach is general and complicates the analysis of fully random ensembles of atoms as well \cite{skip16prb,skip18prl}, its impact is amplified here by the particular narrowness of the frequency range in which the localization transition takes place and the low DOS in this range. Indeed, for scalar waves in a random ensemble of point scatterers studied in Ref.\ \onlinecite{skip16prb}, $\ln g_{q=0.05}$ grows from $\ln g_{q=0.05}^{(c)} - 1$ to $\ln g_{q=0.05}^{(c)} + 1$ in a frequency range $\delta\omega/\Gamma_0 \simeq 0.08$ whereas in the photonic crystal studied here the same growth takes place within $\delta\omega/\Gamma_0 \simeq 0.02$ [see Fig.\ \ref{fig_scaling}(d)]. In addition, DOS of the fully random system has no particular features in the transition region, whereas in the photonic crystal, the localization transition takes place near a band edge where DOS is quite low [see Fig.\ \ref{fig_dos}(b)]. These factors limit the statistical accuracy of our numerical data and make the high-precision estimation of $\nu$ a heavy computational task.

\begin{figure}[t]
\includegraphics[width=\columnwidth]{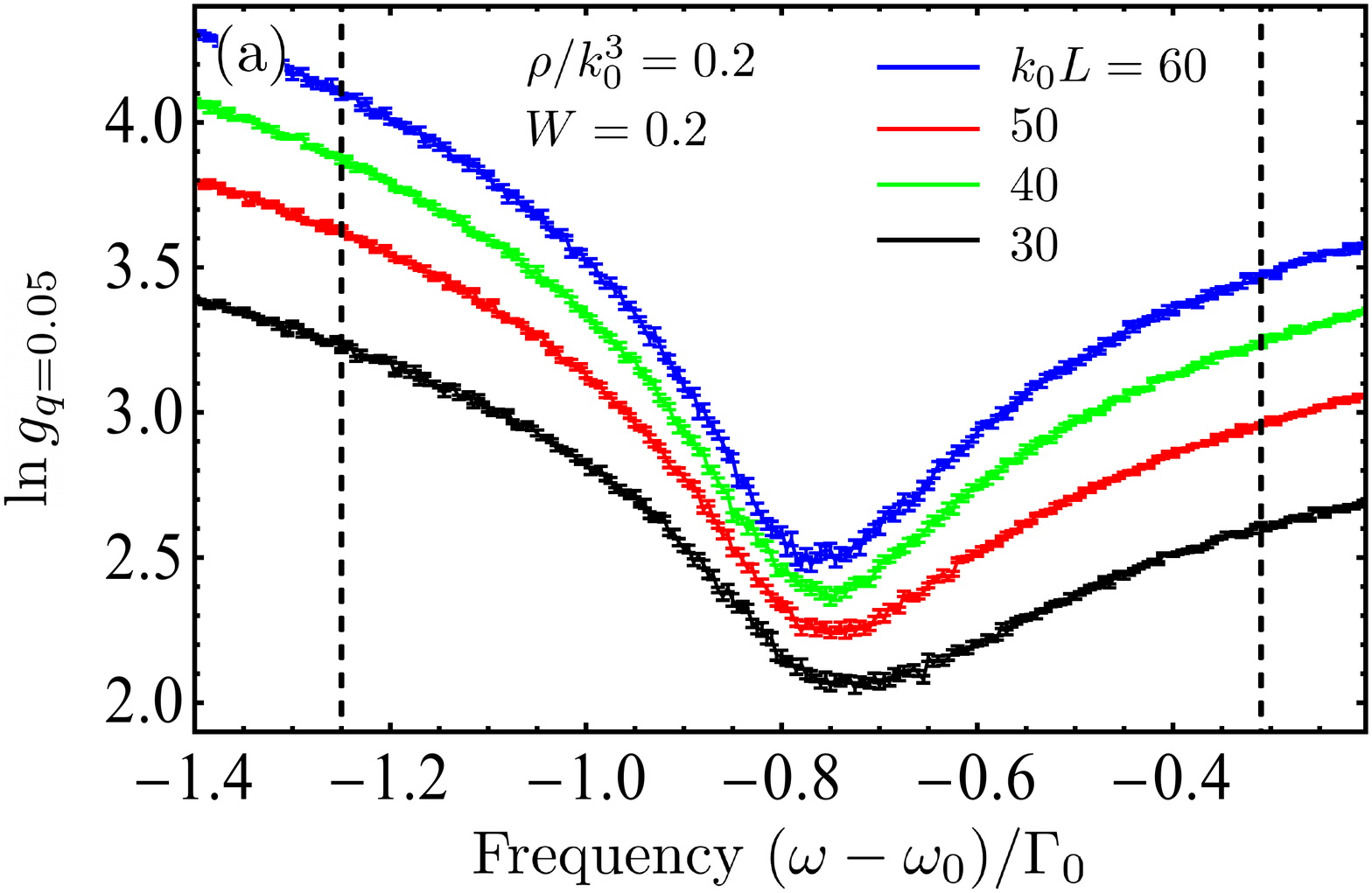}\\
\vspace*{-5mm}
\includegraphics[width=\columnwidth]{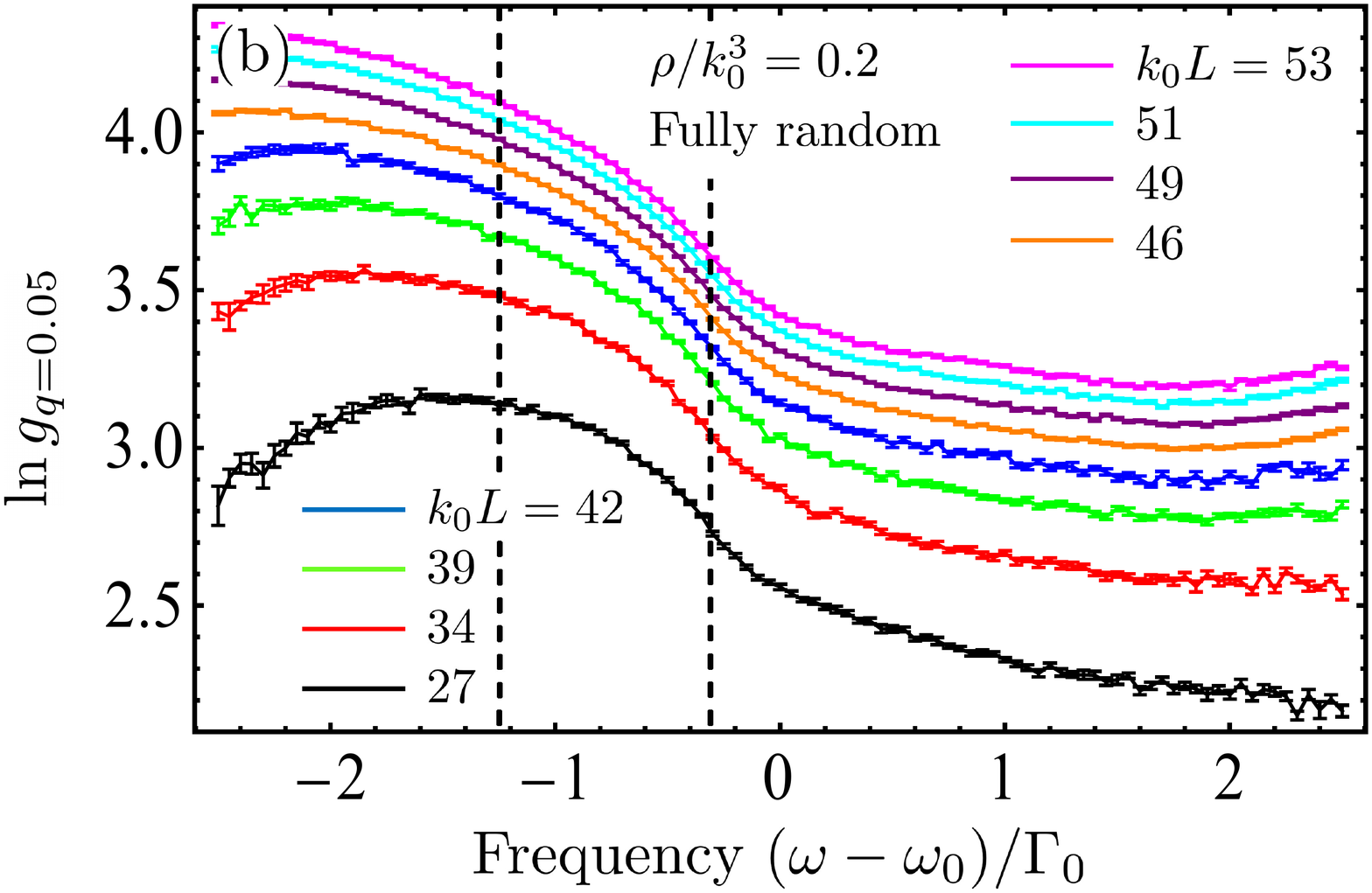}
\vspace{-10mm}
\caption{\label{fig_scaling_random}
Firth percentile $\ln g_{q = 0.05}$ of the Thouless conductance as a function of frequency $\omega$ for different diameters $L$ of a disordered crystal with disorder strength $W = 0.2$ (a) and a fully disordered spherical ensemble of resonant atoms (b). The average number density of atoms is the same as in the photonic crystal analyzed in Fig.\ \ref{fig_scaling}. Vertical dashed lines show band edges of the ideal crystal. The absence of crossings between curves corresponding to different $L$ confirms the absence of localization transitions in these systems.}
\end{figure}

The frequency range in which the quasimodes are localized can be broadened and DOS in this range can be raised by increasing the strength of disorder $W$. However, the space for increase of $W$ without closing the gap and loosing localization altogether is rather limited. As we show in Fig.\ \ref{fig_dos}, the gap closes already for $W = 0.2$, and this closing is accompanied by the disappearance of states localized due to disorder. We illustrate this in Fig.\ \ref{fig_scaling_random}(a) where the firth percentile of conductance is shown as a function of frequency for $W = 0.2$ and the same sizes $L$ of the disordered photonic crystal as in Fig.\ \ref{fig_scaling}. Contrary to the latter figure, no crossings between lines $\ln g_q(\omega,L)$ occur in Fig.\ \ref{fig_scaling_random}(a), signaling the absence of localization transitions. Moreover, the values of $\ln g_q(\omega,L)$ in Fig.\ \ref{fig_scaling_random}(a) are rather high: $\ln g_q(\omega,L) > 2$ for all $\omega$. This means that at any frequency, less than 5\% of $g$ values obtained for different atomic configurations are smaller than $\exp(2) \approx 7$, which is incompatible even with the ``weakest'' form of the Thouless localization criterion requiring that some typical value of $g$ ($\langle g \rangle$, $\exp(\langle g \rangle)$, median $g$, etc.) is less than 1. Finally, another signature of the absence of localization is the monotonous growth of $\ln g_q$ with $L$ at all frequencies indicating that most probably, $\ln g_q \to \infty$ in the limit of $L \to \infty$, as it should be for spatially extended modes.

Further increase of the strength of disorder $W$ beyond $W = 0.2$ does not modify the situation qualitatively as the behavior of the system gets closer to that of a fully random ensemble of atoms studied previously \cite{skip14prl}. The fully random limit is illustrated in Fig.\ \ref{fig_scaling_random}(b) that exhibits the same characteristic features as Fig.\ \ref{fig_scaling_random}(a) (absence of crossings between different curves, large values of $\ln g_q$ and its monotonous growth with $L$) and hence confirms the previously discovered absence of the localization of light in the fully random system \cite{skip14prl}.

The presence of localization only at weak disorder highlights the important differences between localization phenomena in disordered crystals and fully random media. As it has been largely discussed in the literature starting from the pioneering works of Sajeev John \cite{john87,john91,john93}, the localization in a photonic crystal takes place due to an \textit{interplay} of order and disorder in contrast to the localization in a fully random medium that is due to disorder only. Whereas localized states may appear in a \red{3D} random medium only when the strength of disorder exceeds some critical value, even a weak disorder introduces spatially localized modes in the band gap of a disordered photonic crystal and the notion of critical disorder does not exist. However, the possibility of reaching localization at arbitrary weak disorder is counterbalanced by the narrowness of frequency ranges inside the band gap in which the density of states is large enough to allow for observation of the localization of light in an experiment or a numerical simulation. Increasing disorder widens the relevant frequency ranges but also tends to close the band gap and hence to suppress the `order' part of the interplay between order and disorder. A compromise is reached at some intermediate disorder strength that is sufficient to significantly affect wave propagation at frequencies near band edges but not large enough to close the band gap. For the atomic crystal considered in this work, such a compromise seems to be reached around $W = 0.1$ for which the band gap remains open (see Fig.\ \ref{fig_dos}) while localized states become visible (see Fig.\ \ref{fig_ipr_perfect}).

The disappearance of localized modes with the increase of disorder strength $W$ allows for an additional insight about the reasons behind the absence of Anderson localization of light in a completely random 3D ensemble of point scatterers. Indeed, recent work \cite{skip19prb,sgrignuoli20prb} has confirmed the initial suggestion \cite{skip14prl} that the resonant dipole-dipole coupling between scatterers impedes the formation of spatially localized optical modes in 3D. This explanation seems to be supported by the fact that localized modes do arise in a photonic crystal where the distance $\Delta r$ between any two scatterers (atoms) is always larger than a certain minimal distance ($a\sqrt{3}/4$ for a diamond crystal with a lattice constant $a$ considered in this work) and hence the strength of the dipole-dipole coupling between scatterers that scales as $1/\Delta r^3$, is bounded. The increase of $W$ enhances chances for two atoms to be closer, the minimum possible distance between atoms being equal to $(\sqrt{3}/4 - 2W)a$ in our model. The probability for two neighboring atoms to get infinitely close because of disorder becomes different from zero for $W \geq \sqrt{3}/8 \simeq 0.22$. This estimation of disorder strength for which dipole-dipole interactions should become particularly strong, is reasonably close to the approximate value $W \simeq 0.2$ for which localized modes disappear [see Fig.\ \ref{fig_scaling_random}(a)] and the band gap closes (see Fig.\ \ref{fig_dos}). The closeness of the two values suggests a relation between the near-field dipole-dipole interactions, the photonic band gaps, and the spatial localization of optical modes although the exact nature of this relation still remains to be established. Although our analysis supports the arguments based on Eq.\ (\ref{loccrit}) and suggesting that the underlying crystalline structure facilitates the localization phenomenon due to the suppression of DOS near band edges, it also highlights the importance of yet another feature of a crystal---the existence of a minimal distance between two scattering units (atoms or, more generally, ``particles''). At the same time, the impact of the crystalline structure of the atomic lattice on the spatial localization of optical modes does not reduce to keeping neighboring atoms far apart from each other. One of the consequences of the crystalline structure is the fact that in our photonic crystal, the localized modes are red-shifted with respect to the atomic resonance frequency ($\omega < \omega_0$) in contrast to the blue-shifted localized modes that arise in a fully disordered ensemble of atoms in a strong magnetic field \cite{skip15prl,skip18prl}.

A final remark concerns the spatially extended quasimodes in the middle of the band gap, corresponding to large $\ln g_{q=0.05} \gtrsim 2$ between $(\omega-\omega_0)/\Gamma_0 \simeq -0.97$ and $-0.57$ in Fig.\ \ref{fig_scaling}. As we have illustrated already in Fig.\ \ref{fig_ipr_cm}, most of these quasimodes are bound to the surface of the crystal. Their statistical weight is thus expected to decrease with $L$ roughly as the surface-to-volume ratio $1/L$, which tends to zero when $L \to \infty$ but remains significant in our calculations restricted to rather small $L$.  Nevertheless, we clearly see from Figs.\ \ref{fig_scaling}(a--c) that the frequency range in the middle of the bandgap where $\ln g_{q=0.05}$ takes large values $\ln g_{q=0.05} \gtrsim 2$ and remains a globally growing function of $L$, shrinks as $L$ increases. No transition point where curves $\ln g_q$ corresponding to different $L$ cross can be identified around $(\omega-\omega_0)/\Gamma_0 \simeq -0.97$ or $-0.57$, which is especially clear in Fig.\ \ref{fig_scaling}(b) whereas less obvious in Fig.\ \ref{fig_scaling}(c) due to much stronger fluctuations of the numerical data. We note that the above picture of surface modes playing less and less important role as $L$ increases is certainly only a rough approximation to the complete explanation of the evolution of the spectrum in the middle of the band gap. Nontrivial features that are already seen from our results and call for explanation include the nonmonotonous behavior of $\ln g_q$ with $L$ near the high-frequency end of the interval $-0.97 \lesssim (\omega-\omega_0)/\Gamma_0 \lesssim -0.57$ [note the red line that crosses the green line around $(\omega-\omega_0)/\Gamma_0 \simeq -0.7$ in Fig.\ \ref{fig_scaling}(a)] and much stronger fluctuations around $(\omega-\omega_0)/\Gamma_0 \simeq -0.57$ than around $(\omega-\omega_0)/\Gamma_0 \simeq -0.97$ [compare Figs.\ \ref{fig_scaling}(c) and (b)]. Unfortunately, a study of these puzzling features is difficult to perform using our numerical method because it mobilizes significant computational power to obtain the full spectrum of the system of which only a very small fraction [i.e., a small number of eigenvalues $\Lambda_m$ of the matrix (\ref{green})] fall in the band gap where the density of states is low.

\section{Conclusions}
\label{sec:concl}

We performed a thorough theoretical study of the localization of light in a 3D disordered photonic crystal made of two-level atoms. The atoms are first arranged in a diamond lattice with a lattice constant $a$ and are then slightly displaced in random directions by random distances up to $Wa$. We show that spatially localized quasimodes appear near edges of the band gap of the ideal crystal when the disorder strength is $W = 0.1$ or smaller. $W = 0.2$ or larger leads to the closing of the band gap and the disappearance of localized states. The finite-size scaling analysis of the transition between extended and localized states near the high-frequency edge of the band gap suggests that the critical (localization-length) exponent of the transition $\nu$ is in the interval 0.8--1.1, which is different from $\nu_{\text{AM}} \simeq 1.57$ corresponding to the Anderson transition of the 3D orthogonal universality class to which the investigated transition might be expected to belong because of the absence of any particular symmetry breaking mechanisms and, in particular, the preserved time-reversal symmetry.

From the practical standpoint, arranging atoms in a diamond lattice may be a realistic alternative to subjecting them to strong magnetic fields in order to reach the localization of light in cold-atom systems. Indeed, atomic lattices can be readily designed by loading atoms in optical potentials created by interfering laser beams with carefully adjusted phases and propagation directions \cite{greiner02,anderlini07}. Some degree of disorder may arise in such lattices due to experimental imperfections; ways to create additional, controlled disorder have been largely explored in recent years \cite{sanchez10}. Calculations presented in this work provide quantitative estimates for disorder strengths and frequency ranges for which localized quasimodes should appear in lattices of cold atoms featuring a $J_g = 0 \to J_e = 1$ transition. Examples of appropriate chemical elements for vapors of which laser cooling technologies are readily available include strontium (Sr) or ytterbium (Yb). Multiple scattering of light in large ensembles of Sr atoms has been already reported \cite{bidel02} and high atomic number densities have been reached in experiments with Yb \cite{takasu03}. In addition, some of our conclusions may hold for atomic species with more complicated level structure, which may be easier to manipulate and control in an experiment (e.g., rubidium). This opens a way towards the experimental observation of phenomena reported in this work.

\vspace*{-4mm}
\begin{acknowledgments}
\vspace*{-3mm}
All the computations presented in this paper were performed using the Froggy platform of the CIMENT infrastructure (\href{https://ciment.ujf-grenoble.fr}{{\tt https://ciment.ujf-grenoble.fr}}), which is supported by the Rhone-Alpes region (grant CPER07\verb!_!13 CIRA) and the Equip@Meso project (reference ANR-10-EQPX-29-01) of the program {\it Investissements d'Avenir} supervised by the {\it Agence Nationale de la Recherche}.

\end{acknowledgments}

\vspace*{-5mm}

\bibliography{myreferences}

\end{document}